\documentclass[sigconf]{acmart}

\usepackage{graphicx} 
\usepackage{listings}

\setcopyright{none}
\acmYear{2026}
\acmDOI{}
\acmISBN{}
\acmConference[LOCO 2026]{2nd International Workshop on Low Carbon Computing}{September 10--11,
  2026}{Lancaster, UK}

\title{A Scoping Review of Methods to Measure the Energy and Carbon Footprint of Web Tracking and Advertising}

\author{Nils Bonfils}
\email{nils.bonfils@mail.utoronto.ca}
\affiliation{%
  \institution{University of Toronto}
  \city{Toronto}
  \state{Ontario}
  \country{Canada}
}
\author{Christoph Becker}
\email{christoph.becker@utoronto.ca}
\affiliation{%
  \institution{University of Toronto}
  \city{Toronto}
  \state{Ontario}
  \country{Canada}
}

\begin{abstract}
The environmental impact of web tracking and advertising is increasingly receiving attention as the ICT sector's carbon footprint keeps rising. Yet the scholarship addressing this question remains scattered across disciplines and inconsistent in its terminology. This paper presents a scoping review of the literature on methods for measuring the energy and carbon footprint of web tracking and advertising. From an initial pool of 46 articles identified through a structured title-based search on Google Scholar, we arrived at a final corpus of 15 papers, from which we identified five distinct methodological approaches: ad blocking, controlled environment, replaying ads, traffic flow analysis, and literature-derived estimation. This review provides a structured overview of the current methodological landscape and a foundation for more comprehensive environmental accounting of the ad tech ecosystem.
\end{abstract}

\begin{document}

\maketitle

\section{Introduction}

Digital advertising is the growing primary source of revenue for the majority of services offered on the web, including search engines, social media platforms, news outlets, and video streaming platforms \cite{pwcIABInternetAdvertising2026}. Those services represent the core products of multibillion-dollar corporations that derive the majority of their revenue through advertising \cite{alphabetAlphabetInvestorRelations,metaMetaFinancials}. To increase the relevance of digital advertisements, web advertising is often coupled with user tracking (a.k.a. analytics or behavioral/targeted advertising) forming part of a complex "ad tech" ecosystem \cite{lomborgDigitalTrackingInfrastructural2023}.

However, tracking and advertising on the web extends far beyond traditional advertising. They represent the user-facing layer of a complex ecosystem of software and data transfer engineered to generate profit from users' interactions with ostensibly free web services. Sometimes called surveillance capitalism \cite{zuboffAgeSurveillanceCapitalism2019}, digital capitalism \cite{schillerDigitalCapitalismNetworking2000}, or platform capitalism \cite{srnicekPlatformCapitalism2017}, this new economic order is built on the monetization of enormous quantities of user data. The infrastructures required to capture, store, and process data at scale consumes substantial energy and carries significant material consequences. Accurately calculating the environmental cost of surveillance capitalism practices is therefore a necessary step toward a fuller picture of its ecological toll, of which web tracking and advertising is the most visible constituent part.

While generic software tools to measure the carbon footprint of software and services exist, accurately measuring the energy and carbon cost of web tracking and advertising poses a distinct methodological challenge: isolating the computational resources specifically devoted to displaying advertisements and capturing user data from the resources required by the regular operation of the software. Existing research does address the environmental impact of web tracking and advertising, but it draws on a diversity of approaches without any coherent understanding of what the state-of-the-art methods are or how they relate to one another. Addressing that gap is what this paper seeks to accomplish: \textit{what methods have been established to quantify the computational resource consumption of web tracking and advertising?} To that end, we conduct a scoping review of the literature on methods to measure the energy and carbon footprint of web tracking and advertising.

\section{Motivation}



The impetus for this work stems from the authors' earlier efforts to map and assess the environmental impact of surveillance capitalism \cite{bonfilsEmpiricalInquirySurveillance2025d,bonfilsEnvironmentalCostsSurveillance2026a}. In that prior work, the authors sought methods to calculate the environmental impact of processes closely related to web advertising and analytics, but found no clear answer in the existing literature. The literature quantifying the carbon and energy footprint of ad tech does exist but is scattered across different fields, such as computer science, impact assessment, and industrial ecology. This disciplinary fragmentation has produced considerable terminological inconsistency: terms like web analytics, online tracking, and user tracking typically denote the same processes, while words like costs, impacts, and footprint are used interchangeably across studies. Combined with the relative recency of this phenomenon, this terminological disjunction makes it difficult to follow the latest methodological developments in the field.

Nevertheless, a substantial body of research has recently emerged addressing these questions. This body of literature is comprised of studies that draw on a variety of methods involving bespoke setups for data collection, diverse metrics, and differing statistical methods. Given the range of studies and the absence of a unified field or community regrouping the knowledge surrounding it, a scoping review is an appropriate instrument for providing an initial overview of the methods used to measure the energy and carbon footprint of web tracking and advertising.

\section{Methods}


Our approach started with a base set of ten papers, each highly relevant to measuring the energy and carbon impact of advertising, tracking, and surveillance on the web (Appendix \ref{sec:initialpapers}). The ten base papers were identified through a non-systematic combination of online and database searches, including DuckDuckGo, Google Scholar, ACM Digital Library, IEEE Xplore, alongside both backward and forward snowballing. From this set, we extracted terms from the paper titles to capture both the topic of focus and the type of impact measured (Appendix \ref{sec:queryterms}), which were then used to craft search queries. 

For this review, we selected Google Scholar as our search database. We restricted our searches to paper titles by using the "allintitle:" operator, and excluded a small number of terms to filter out unrelated papers. It is worth noting that Google Scholar's query syntax is both limited and not well-documented. Throughout our searches, we discovered a 340 character ceiling beyond which query terms are silently dropped while still appearing in the search box. In order to circumvent that limitation, we divided our search into multiple queries that are detailed in Appendix \ref{sec:finalqueries}. This process ended up yielding a pool of 46 articles.

We then filtered this pool based on the title, abstract, and a brief review of the methods section, retaining only papers that attempted to measure the energy or carbon footprint of web tracking, advertising, or associated network traffic. This filtered set contained 18 articles, from which we excluded three: 1) a thesis \cite{khanImpactAdBlockers2025}, 2) an online news article \cite{oakesCarbonFootprintDigital2021}, and 3) a conference paper that we were unable to access \cite{visserEffectAdBlockers2016}. This left a final corpus of 15 papers (Table \ref{tab:papers}).


\begin{table*}[htbp]
    \centering
    \caption{The final set of 15 papers. Each work uses either a time-based or network traffic-based metric. Each work focus on different part of the ecosystem of ad and tracking. Each work makes use of one of the five methods discussed in the findings.}
    \begin{tabular}{ | c | c | c | c | }
        \hline
        \textbf{Paper} & \textbf{Metric} & \textbf{Focus} & \textbf{Method} \\
        \hline\hline
        (Simons and Pras, 2010)\cite{simonsHiddenEnergyCost2010} & Time-based & Client-side & Ad Blocking \\
        \hline
        (Pärssinen et al., 2018)\cite{parssinenEnvironmentalImpactAssessment2018} & Network Traffic & Holistic & Literature-Derived Estimation \\
        \hline
        (Pearce, 2020)\cite{pearceEnergyConservationOpen2020} & Time-based & Client-side & Ad Blocking \\
        \hline
        (Cuccietti et al., 2022)\cite{cucchiettiCarbolyticsAnalysisCarbon2022} & Network Traffic & Holistic & Traffic Flow Analysis \\
        \hline
        (González-Cabañas et al., 2023)\cite{gonzalez-cabanasCarbonTagBrowserBasedMethod2023} & Time-based & Client-side & Replaying Ads \\
        \hline
        (Pesari et al., 2023)\cite{pesariClientsideEnergyGHGs2023} & Time-based & Client-side & Ad Blocking \\
        \hline
        (Khan et al., 2024a)\cite{khanPowerConsumptionUsing2024} & Time-based & Client-side & Ad Blocking \\
        \hline
        (Khan et al., 2024b)\cite{khanImpactAdBlockers2024} & Time-based & Client-side & Ad Blocking \\
        \hline
        (Khan et al., 2024c)\cite{khanImpactBuiltinAdBlockers2024} & Time-based & Client-side & Ad Blocking \\
        \hline
        (Petalotis et al., 2024)\cite{petalotisEmpiricalStudyPerformance2024} & Time-based & Client-side & Controlled Environment \\
        \hline
        (Puhtila et al., 2024)\cite{puhtilaEffectAnalyticsTools2024} & Time-based & Client-side & Controlled Environment \\
        \hline
        (Khan et al., 2025)\cite{khanComputerPowerConsumption2025} & Time-based & Client-side & Ad Blocking \\
        \hline
        (Duprat, 2026)\cite{dupratThermodynamicEfficiencyInversion2026} & Network Traffic & Client and Server & Literature-Derived Estimation \\
        \hline
        (Pélissier et al., 2026)\cite{pelissierUsersPayTwice2026} & Time-based & Client-side & Ad Blocking \\
        \hline
        (Bonfils et al., 2026)\cite{bonfilsEnvironmentalCostsSurveillance2026a} & Network Traffic & Connectivity & Traffic Flow Analysis \\
        \hline
    \end{tabular}
    \label{tab:papers}
\end{table*}

\section{Findings}

During our analysis of the articles, we found it useful to consider several of the dimensions outlined by Pärssinen et al. \cite{parssinenEnvironmentalImpactAssessment2018}. The first dimension concerns how contributions to the carbon impact of the ICT sector are allocated, based on either the use time or by the volume of data traffic. The second dimension is the fundamental internet subsystem addressed by the paper. It can be one of: 1) the end-user devices: \textit{client-side}, 2) the infrastructure connecting users to applications: \textit{connectivity}, 3) the applications themselves: \textit{server-side}, or 4) the traffic flowing across other sub-systems: \textit{secondary traffic}. The third dimension concerns the approach used to estimate the power usage or carbon footprint of a system, typically one of: 1) top-down, 2) bottom-up, 3) model based, or 4) unified method. Finally, we identified five distinct methods across all the articles in our final corpus listed in Table \ref{tab:papers}.

\subsection{Method 1: Ad Blocking}
\label{sec:adblocking}
The most common approach to measuring the carbon and energy footprint of advertising and analytics in our selected scholarship involves browsing websites that contains ads, with and without some form of ad blocking. Ad blocking most commonly takes the form of a web browser extension, such as uBlock Origin, Adblock Plus, or AdGuard. However, ad blocking can also be implemented as a built-in feature of web browsers (i.e. Brave, Opera, Vivaldi)\cite{khanImpactBuiltinAdBlockers2024}, filtered through an HTTP proxy \cite{simonsHiddenEnergyCost2010}, or even blocked at the DNS request level\cite{pelissierUsersPayTwice2026}.

For clarity, we deconstruct this approach into three essential steps:
\begin{enumerate}
    \item Define a navigation or web browsing strategy
    \item Select ad blocking configurations
    \item Instrument the measurement setup
\end{enumerate}

\paragraph{Step 1: Navigation Strategy.}
The first step of the ad blocking approach involves defining which websites will be browsed and how. Most studies select a sample from online lists of popular websites (e.g. Digg, Tranco, or 4imn.com), some (such as Khan et al.) simply use an arbitrary list of websites \cite{khanComputerPowerConsumption2025,khanImpactAdBlockers2024,khanImpactBuiltinAdBlockers2024,khanPowerConsumptionUsing2024}. Methodological transparency is achieved by clearly documenting the initial source list, the inclusion and exclusion criteria, and the motivation behind those choices.

Once the list of websites to be browsed is finalized, a browsing approach must be devised. Some studies instrument a web browser to automate browsing using tools, such as Puppeteer \cite{pelissierUsersPayTwice2026} or AutoBrowse, a bespoke solution built specifically by Simons and Pras for their study \cite{simonsHiddenEnergyCost2010}. The browsing behavior can range from simply loading a page, to automating mouse movement \cite{pesariClientsideEnergyGHGs2023}, to emulating realistic user behavior \cite{pelissierUsersPayTwice2026}.

\paragraph{Step 2: Ad Blocking Configuration.}
The ad blocking configuration plays an important role, as most studies demonstrated a significant variation in the power measured depending on the combination of web browser, type of ad blocking (browser extension, built-in, proxy, or DNS), OS, and hardware. Configurations are therefore always disclosed, with Khan et al. going as far as to organize their works around specific categories of configurations \cite{khanComputerPowerConsumption2025,khanImpactAdBlockers2024,khanImpactBuiltinAdBlockers2024,khanPowerConsumptionUsing2024}.

\paragraph{Step 3: Measurement Setup.}
The measurement setup can take several different forms. Simons and Pras opted for direct measurements with a voltmeter to record hardware power usage \cite{simonsHiddenEnergyCost2010}. Pearce recorded differences in page load time to infer how much time was "lost" to advertisement delivery and tracking \cite{pearceEnergyConservationOpen2020}. Most other studies use a software-based approach that draws on CPU or battery information to infer power consumption -- common tools for this include Powerstat, HWiNFO, and Batterystat.

\subsection{Method 2: Controlled Environment}
\label{sec:controlledenv}
Another approach, similar in essence to ad blocking but different in execution, is what we term the "controlled environment" method. Rather than blocking ads after the fact, this method involves creating or recreating websites in lab settings in order to control the delivery side of ads and trackers. The two studies that employed this method implemented it in notably different ways. 

Petalotis et al. followed a similar approach to the ad blocking method, selecting a list of nine websites containing both ads and analytics and automating web browsing on Android using MonkeyRunner. However, rather than configuring ad blocking, they downloaded the selected websites and produced three versions of each: 1) the original, unchanged; 2) with ads removed; and 3) with analytics and trackers removed. These versions were then served locally in a controlled environment, where measurements were carried out \cite{petalotisEmpiricalStudyPerformance2024}.

Puhtila et al. implemented this approach differently, building a website from scratch and instrumenting a varying number of analytical tools (0, 1, 5 and 10) on that website. Their navigation strategy used an automated setup with Selenium to perform actions on the website, including page navigation, scrolling on a page, and form submission, while power consumption was measured directly via voltmeter \cite{puhtilaEffectAnalyticsTools2024}.

\subsection{Method 3: Replaying Ads}
This approach is used by a single paper \cite{gonzalez-cabanasCarbonTagBrowserBasedMethod2023}. González-Cabañas et al. first created a dataset by automatically crawling a thousand popular websites and saving the ads encountered, gathering around 25,000 ads. In parallel, they manually collected around 500 ads from humans browsing the web. They then built an automated setup to render and display each collected ad individually. The measurement process ran in two steps: first, opening and closing a blank page of the browser to establish a "baseline" measurement; second, opening the ad in the web browser and recording its energy usage. Subtracting the baseline from the second measurement yielded the energy cost of rendering a single ad. To generalize those findings, they trained a machine learning model on the pairing of ads with their normalized energy usage, demonstrating how statistical models can extend the reach of such a method.


\subsection{Method 4: Traffic Flow Analysis}
\label{sec:trafficflowanalysis}
This method consists in analyzing the network traffic generated by advertisement and tracking processes. Cuccietti et al. \cite{cucchiettiCarbolyticsAnalysisCarbon2022} used OpenWPM, a tool that extends and optimizes Selenium for crawling large number of websites \cite{englehardtOnlineTracking1millionsite2016a}, to crawl the top one million websites according to the Tranco List and capture the network traffic for each page load. They used cookies as the differentiator between tracking and non-tracking traffic, enabling them to leverage the network traffic data collected to sum the amount of bytes used by cookies, and estimated the total data transferred for cookie-based tracking. As appealing as deriving energy and carbon footprint estimates from this figure by applying the appropriate energy and carbon intensity coefficients may be, it is important to note that such an approach is flawed and will likely results in misleading figures. Current work is underway attempting to attribute the energy consumption associated with data traffic \cite{schienCausalAllocationFixed2025}.

The authors' most recent work is a pilot study quantifying the environmental footprint of user tracking and other processes related to surveillance capitalism \cite{bonfilsEnvironmentalCostsSurveillance2026a}. The approach compares the network traffic usage on two distinct platforms: a corporate platform whose business model requires the use of ads and analytics (i.e. X/Twitter) and a non-commercial free and open-source counterpart free of those incentives (i.e. Mastodon). The network traffic was measured by instrumenting a web browser to emulate realistic user behavior (i.e. scrolling on the feed and making posts).

\subsection{Method 5: Literature-Derived Estimation}
The final method is the only approach that requires no primary data collection and no dedicated measurement setup. Instead, it derives energy and carbon footprint estimates by drawing on coefficients and figures reported in existing literature, applying them relevant metrics to produce macro-level consumption estimates. Duprat, offers a lightweight application of this method: estimating the proportion of computational resources used by advertisements and applying carbon intensity coefficients to calculate energy and carbon footprint estimates \cite{dupratThermodynamicEfficiencyInversion2026}. This approach is accessible but its validity rests on the relevance of the coefficients it borrows. 

Pärssinen et al. are considerably more thorough \cite{parssinenEnvironmentalImpactAssessment2018}. Drawing on an extensive body of literature, they contribute a framework for assessing the environmental impact of any internet service. The least transparent element of their analysis is their estimate of the share of network traffic dedicated to online advertising, which could not be grounded in prior literature as it was non-existent at the time. Nevertheless, their approach demonstrated clearly how to handle uncertainty in figures derived from the literature.

\subsection{Presentation of Results}

It is worth noting that none of the works directly measure CO2 emissions. Instead, they translate their chosen metric into energy (typically in kWh), if power was not measured directly, and then convert that energy figure to a CO2-equivalent quantity by multiplying it with a recent local or global grid carbon intensity coefficient. Some works contribute a figure representing the relative power consumption of web tracking or advertising without attempting to extrapolate their results to scale. For instance, Khan et al. report that using a web browser with a built-in ad blocker can reduce power consumption by up to 44\% compared to traditional web browser setups \cite{khanImpactBuiltinAdBlockers2024}. Conversely, other works, such as Pearce, go as far as to estimate the financial savings and the number of lives that could be saved from coal-based pollution if all the internet users in the United States used an ad blocker \cite{pearceEnergyConservationOpen2020}.

\section{Discussion}


This scoping review identified five distinct methodological approaches across the 15 studies of our final corpus. Of these, the ad blocking method is adopted by the greatest number of different authors (despite the bias towards Khan et al.), offering a practical and reproducible means of isolating the energy cost of ads and trackers on the client side. The controlled environment and individual ad replay methods offer greater experimental control at the cost of validity, while traffic flow analysis and literature review represent approaches better suited to macro-level estimation. No single method has emerged as a standard, and the field has yet to converge on shared metrics, terminology, and evaluations.

Our corpus suggests that the interest for this area of inquiry is rather young and growing with 12 out of the 15 studies (80\% of the corpus) published in the last five years. This recent increase in attention is understandable, particularly in light of the global environmental crisis and the growing environmental impact attributable to the ICT sector \cite{freitagRealClimateTransformative2021}. The urgency of this topic is further compounded by the critiques of surveillance capitalism, which demonstrate how the economical and political incentives underpinning ad tech are harmful for democratic societies \cite{zuboffAgeSurveillanceCapitalism2019}. The link between the concrete environmental impacts of ad tech and the critiques of surveillance is a novel angle that we, the authors, are seeking to explore. 

Despite its small size, the corpus revealed a breadth of methodological approaches that also reflects fragmentation in the field. The studies engage with shared questions without meaningfully building on or responding to each other's findings. For instance, when González-Cabañas et al. discovered that the transfer size (i.e. the amount of data transferred necessary to receive the ad) is irrelevant to estimate the energy necessary to render individual ads \cite{gonzalez-cabanasCarbonTagBrowserBasedMethod2023}, they did not situate this finding in relation to the work of Pärssinen et al. whose analysis relies on allocating energy consumption based on network traffic \cite{parssinenEnvironmentalImpactAssessment2018}. The implication is significant: if transfer size is not a reliable proxy for the energy consumed by client-side ad rendering, then Pärssinen et al's estimate of the energy consumption of online advertising on user devices may be invalid.

Several directions emerge as priorities for future work. Most immediately, the near-complete absence of server-side and secondary traffic measurements are highlighting the need to develop methods to instrument the measurement of ad delivery and user data processing within data center infrastructure. However, the systemic opacity of corporate infrastructure undermines the feasibly of such undertakings. Thus, the controlled environment approach (Section \ref{sec:controlledenv}) offers an imperfect but accessible approach that warrants further development as a methodology capable of addressing that gap, given its potential to model the full stack. There is also a need for shared terminology and methods across the disciplines for coherent dialogue between studies. This could be achieved through a systematic literature review that extends the queries used in this review, and includes terms such as "telemetry" as a synonym for analytics and "efficiency" as a synonym for consumption, as well as emphasizing the need to capture literature on the server-side. Finally, longitudinal studies tracking the environmental cost of ad tech over time are lacking and are essential to track the evolution and trends of the environmental impact of this industry.

It is important to underscore that while this paper is a scoping review, it is not systematic \cite{munnSystematicReviewScoping2018}. We deliberately limited our search to terms in the title of papers and did not employ complementary methods, such as backward or forward snowballing. Therefore, the five methods presented here should not be considered exhaustive and are instead representative. We also note that this work only engages with literature measuring the footprint of web tracking and advertising. We recognize that a wider scope of would surface more approaches capable of quantifying the energy and carbon footprint of software and web services (e.g. CO2.js\footnote{\url{https://www.thegreenwebfoundation.org/co2-js/}}).

\section{Conclusion}
This scoping review maps existing methods to measure the energy and carbon footprint of web tracking and advertising, a question that sits at the intersection of the environmental crisis and the growing critique of surveillance capitalism. Through a structured literature search and filtering process, we identified 15 relevant studies and surfaced five distinct methodological approaches: ad blocking, controlled environment, replaying ads, traffic flow analysis, and literature-derived estimation. The field itself seems to be gaining momentum with 80\% of the papers in our corpus published within the last five years. This growing interest combined with a variety of existing methods suggests the importance and potential of quantifying the environmental costs of the ad tech ecosystem, while the methodological groundwork for doing so is already taking shape.

\bibliographystyle{ACM-Reference-Format}
\bibliography{refs}

@misc{cucchiettiCarbolyticsAnalysisCarbon2022,
  title = {Carbolytics: {{An Analysis}} of the {{Carbon Costs}} of {{Online Tracking}}.},
  author = {Cucchietti, Fernando and Moll, Joana and Esteban, Marta and Reyes, Patricio and Calatrava, Carlos Garcia},
  year = 2022,
  month = feb,
  urldate = {2025-10-27},
  howpublished = {https://carbolytics.org/report.html},
  langid = {english}
}

@misc{dupratThermodynamicEfficiencyInversion2026,
  type = {{{SSRN Scholarly Paper}}},
  title = {The {{Thermodynamic Efficiency Inversion}}: {{A Comparative Energy Lifecycle Assessment}} of {{Generative AI Inference}} versus {{Ad-Supported Web Search Sessions}}},
  shorttitle = {The {{Thermodynamic Efficiency Inversion}}},
  author = {Duprat, Charles},
  year = 2026,
  month = feb,
  number = {6287918},
  eprint = {6287918},
  publisher = {Social Science Research Network},
  address = {Rochester, NY},
  doi = {10.2139/ssrn.6287918},
  urldate = {2026-05-20},
  archiveprefix = {Social Science Research Network},
  langid = {english}
}

@article{gonzalez-cabanasCarbonTagBrowserBasedMethod2023,
  title = {{{CarbonTag}}: {{A Browser-Based Method}} for {{Approximating Energy Consumption}} of {{Online Ads}}},
  shorttitle = {{{CarbonTag}}},
  author = {{Gonz{\'a}lez-Caba{\~n}as}, Jos{\'e} and Callejo, Patricia and Cuevas, Rub{\'e}n and Svartberg, Steffen and Torjesen, Tommy and Cuevas, {\'A}ngel and Pastor, Antonio and Kotila, Mikko},
  year = 2023,
  month = oct,
  journal = {IEEE Transactions on Sustainable Computing},
  volume = {8},
  number = {4},
  pages = {739--750},
  issn = {2377-3782},
  doi = {10.1109/TSUSC.2023.3286916},
  urldate = {2026-05-20}
}

@article{khanComputerPowerConsumption2025,
  title = {Computer {{Power Consumption}} While Using {{Ad-Blocker}} on a {{System}} with {{AI Accelerators}}},
  author = {Khan, Khan Awais and Iqbal, Mohammad Tariq and Iqbal, Mohsin},
  year = 2025,
  month = jan,
  journal = {European Journal of Information Technologies and Computer Science},
  volume = {5},
  number = {1},
  pages = {11--20},
  issn = {2736-5492},
  doi = {10.24018/compute.2025.5.1.144},
  urldate = {2026-05-20},
  langid = {english}
}

@article{khanImpactAdBlockers2024,
  title = {Impact of {{Ad Blockers}} on {{Computer Power Consumption}} While {{Web Browsing}}: {{A Comparative Analysis}}},
  shorttitle = {Impact of {{Ad Blockers}} on {{Computer Power Consumption}} While {{Web Browsing}}},
  author = {Khan, Khan Awais and Iqbal, Mohammad Tariq and Jamil, Mohsin},
  year = 2024,
  month = oct,
  journal = {European Journal of Electrical Engineering and Computer Science},
  volume = {8},
  number = {5},
  pages = {18--24},
  issn = {2736-5751},
  doi = {10.24018/ejece.2024.8.5.650},
  urldate = {2025-04-19},
  copyright = {Copyright (c) 2024 Khan Awais Khan, Mohammad Tariq Iqbal, Mohsin Jamil},
  langid = {english}
}

@mastersthesis{khanImpactAdBlockers2025,
  title = {Impact of Ad Blockers on Computer Power Consumption: A Comparative Analysis of Browser Ad on and Built-in Browsers Feature},
  shorttitle = {Impact of Ad Blockers on Computer Power Consumption},
  author = {Khan, Khan Awais},
  year = 2025,
  month = feb,
  urldate = {2025-04-19},
  copyright = {thesis\_license},
  langid = {english},
  school = {Memorial University of Newfoundland}
}

@article{khanImpactBuiltinAdBlockers2024,
  title = {The {{Impact}} of {{Built-in Ad-Blockers}} in {{Web Browsers}} on {{Computer Power Consumption}}},
  author = {Khan, Khan Awais and Iqbal, Mohammad Tariq and Jamil, Mohsin},
  year = 2024,
  month = nov,
  journal = {European Journal of Information Technologies and Computer Science},
  volume = {4},
  number = {5},
  pages = {1--10},
  issn = {2736-5492},
  doi = {10.24018/compute.2024.4.5.137},
  urldate = {2026-05-20},
  langid = {english}
}

@inproceedings{khanPowerConsumptionUsing2024,
  title = {Power {{Consumption While Using Ad-Blockers}} on {{ARM-Based CPU}}},
  booktitle = {The 33rd {{Annual Newfoundland Electrical}} and {{Computer Engineering Conference}}},
  author = {Khan, Khan Awais and Iqbal, Mohammad Tariq and Jamil, Mohsin},
  year = 2024,
  langid = {english}
}

@misc{oakesCarbonFootprintDigital2021,
  title = {Carbon Footprint of Digital Ads Laid Bare by {{Good-Loop}} Tool},
  author = {Oakes, Omar},
  year = 2021,
  month = jun,
  journal = {The Media Leader},
  urldate = {2026-05-20},
  chapter = {Digital},
  langid = {british}
}

@article{parssinenEnvironmentalImpactAssessment2018,
  title = {Environmental Impact Assessment of Online Advertising},
  author = {P{\"a}rssinen, M. and Kotila, M. and Cuevas, R. and Phansalkar, A. and Manner, J.},
  year = 2018,
  month = nov,
  journal = {Environmental Impact Assessment Review},
  volume = {73},
  pages = {177--200},
  issn = {0195-9255},
  doi = {10.1016/j.eiar.2018.08.004},
  urldate = {2025-06-06}
}

@article{pearceEnergyConservationOpen2020,
  title = {Energy {{Conservation}} with {{Open Source Ad Blockers}}},
  author = {Pearce, Joshua M.},
  year = 2020,
  month = jun,
  journal = {Technologies},
  volume = {8},
  number = {2},
  pages = {18},
  publisher = {Multidisciplinary Digital Publishing Institute},
  issn = {2227-7080},
  doi = {10.3390/technologies8020018},
  urldate = {2026-05-14},
  copyright = {http://creativecommons.org/licenses/by/3.0/},
  langid = {english}
}

@inproceedings{pelissierUsersPayTwice2026,
  title = {Users {{Pay Twice}}: {{The Hidden Energy Cost}} of {{Web Advertising}}},
  shorttitle = {Users {{Pay Twice}}},
  booktitle = {Proceedings of the {{ACM Web Conference}} 2026},
  author = {P{\'e}lissier, Samuel and Mehanna, Naif and Roux, Sterenn and Perez, Quentin and Rudametkin, Walter and Bourcier, Johann and Laperdrix, Pierre},
  year = 2026,
  month = apr,
  series = {{{WWW}} '26},
  pages = {1629--1639},
  publisher = {Association for Computing Machinery},
  address = {New York, NY, USA},
  doi = {10.1145/3774904.3792414},
  urldate = {2026-05-07},
  isbn = {979-8-4007-2307-0}
}

@article{pesariClientsideEnergyGHGs2023,
  title = {Client-Side Energy and {{GHGs}} Assessment of Advertising and Tracking in the News Websites},
  author = {Pesari, Fabio and Lagioia, Giovanni and Paiano, Annarita},
  year = 2023,
  journal = {Journal of Industrial Ecology},
  volume = {27},
  number = {2},
  pages = {548--561},
  issn = {1530-9290},
  doi = {10.1111/jiec.13376},
  urldate = {2025-06-06},
  copyright = {\copyright{} 2022 by the International Society for Industrial Ecology.},
  langid = {english}
}

@article{petalotisEmpiricalStudyPerformance2024,
  title = {An Empirical Study on the Performance and Energy Costs of Ads and Analytics in Mobile Web Apps},
  author = {Petalotis, Christos and Krumpak, Luka and Floroiu, Maximilian Stefan and Ahmad, Lar{\'e}b Fatima and Athreya, Shashank and Malavolta, Ivano},
  year = 2024,
  month = feb,
  journal = {Information and Software Technology},
  volume = {166},
  pages = {107370},
  issn = {0950-5849},
  doi = {10.1016/j.infsof.2023.107370},
  urldate = {2025-06-06}
}

@inproceedings{puhtilaEffectAnalyticsTools2024,
  title = {The {{Effect}} of {{Analytics Tools}} on {{Energy Consumption}} of {{Websites}}},
  booktitle = {2024 10th {{International Conference}} on {{ICT}} for {{Sustainability}} ({{ICT4S}})},
  author = {Puhtila, Panu and Kivim{\"a}ki, Lauri and Heino, Timi and M{\"a}kel{\"a}, Jari-Matti and Rauti, Sampsa and M{\"a}kil{\"a}, Tuomas},
  year = 2024,
  month = jun,
  pages = {335--345},
  doi = {10.1109/ICT4S64576.2024.00041},
  urldate = {2026-01-15}
}

@article{simonsHiddenEnergyCost2010,
  title = {{The Hidden Energy Cost of Web Advertising}},
  author = {Simons, R. J. G. and Pras, Aiko},
  year = 2010,
  month = jun,
  publisher = {{Centre for Telematics and Information Technology (CTIT)}},
  urldate = {2025-06-06},
  langid = {Undefined}
}

@inproceedings{visserEffectAdBlockers2016,
  title = {The {{Effect}} of {{Ad Blockers}} on the {{Energy Consumption}} of {{Mobile Web Browsing}}},
  author = {Visser, A.},
  year = 2016,
  urldate = {2026-05-20}
}

@article{munnSystematicReviewScoping2018,
  title = {Systematic Review or Scoping Review? {{Guidance}} for Authors When Choosing between a Systematic or Scoping Review Approach},
  shorttitle = {Systematic Review or Scoping Review?},
  author = {Munn, Zachary and Peters, Micah D. J. and Stern, Cindy and Tufanaru, Catalin and McArthur, Alexa and Aromataris, Edoardo},
  year = 2018,
  month = nov,
  journal = {BMC Medical Research Methodology},
  volume = {18},
  number = {1},
  pages = {143},
  issn = {1471-2288},
  doi = {10.1186/s12874-018-0611-x},
  urldate = {2026-05-15},
  langid = {english}
}

@book{schillerDigitalCapitalismNetworking2000,
  title = {Digital {{Capitalism}}: {{Networking}} the {{Global Market System}}},
  shorttitle = {Digital {{Capitalism}}},
  author = {Schiller, Daniel},
  year = 2000,
  month = feb,
  publisher = {MIT Press},
  address = {Cambridge, MA, USA},
  isbn = {978-0-262-69233-5},
  langid = {english}
}

@book{zuboffAgeSurveillanceCapitalism2019,
  title = {The {{Age}} of {{Surveillance Capitalism}}: {{The Fight}} for a {{Human Future}} at the {{New Frontier}} of {{Power}}},
  shorttitle = {The {{Age}} of {{Surveillance Capitalism}}},
  author = {Zuboff, Shoshana},
  year = 2019,
  month = jan,
  publisher = {PublicAffairs},
  googlebooks = {lRqrDQAAQBAJ},
  isbn = {978-1-61039-570-0},
  langid = {english}
}

@book{srnicekPlatformCapitalism2017,
  title = {Platform {{Capitalism}}},
  author = {Srnicek, Nick},
  year = 2017,
  month = may,
  publisher = {John Wiley \& Sons},
  googlebooks = {2HdNDwAAQBAJ},
  isbn = {978-1-5095-0488-6},
  langid = {english}
}

@inproceedings{bonfilsEmpiricalInquirySurveillance2025d,
  title = {An {{Empirical Inquiry}} into {{Surveillance Capitalism}}: {{Web Tracking}}},
  booktitle = {11th {{Workshop}} on {{Computing}} within {{Limits}}},
  author = {Bonfils, Nils},
  year = 2025,
  month = jun,
  doi = {10.48550/arXiv.2508.07454},
  langid = {english}
}

@inproceedings{bonfilsEnvironmentalCostsSurveillance2026a,
  title = {The {{Environmental Costs}} of {{Surveillance Capitalism}}: {{A Case Study}} of {{Social Media Platforms}}},
  booktitle = {Accepted at the 12th {{International Conference}} on {{ICT}} for {{Sustainability}}},
  author = {Bonfils, Nils and Becker, Christoph},
  year = 2026,
  month = jun,
  address = {Bern, Switzerland},
  doi = {10.48550/arXiv.2605.26314}
}

@inproceedings{englehardtOnlineTracking1millionsite2016a,
  title = {Online {{Tracking}}: {{A}} 1-Million-Site {{Measurement}} and {{Analysis}}},
  shorttitle = {Online {{Tracking}}},
  booktitle = {Proceedings of the 2016 {{ACM SIGSAC Conference}} on {{Computer}} and {{Communications Security}}},
  author = {Englehardt, Steven and Narayanan, Arvind},
  year = 2016,
  month = oct,
  pages = {1388--1401},
  publisher = {ACM},
  address = {Vienna Austria},
  doi = {10.1145/2976749.2978313},
  urldate = {2026-01-09},
  isbn = {978-1-4503-4139-4},
  langid = {english}
}

@article{schienCausalAllocationFixed2025,
  title = {Causal Allocation of Fixed Impacts in Product Systems: {{Assessing}} the Effect of Data Demand on Network Energy Consumption},
  shorttitle = {Causal Allocation of Fixed Impacts in Product Systems},
  author = {Schien, Daniel and Shabajee, Paul and Krug, Louise and McSorley, Greg and Preist, Chris},
  year = 2025,
  journal = {Journal of Industrial Ecology},
  volume = {29},
  number = {5},
  pages = {1618--1631},
  issn = {1530-9290},
  doi = {10.1111/jiec.70057},
  urldate = {2026-06-13},
  copyright = {\copyright{} 2025 The Author(s). Journal of Industrial Ecology published by Wiley Periodicals LLC on behalf of International Society for Industrial Ecology.},
  langid = {english}
}

@online{alphabetAlphabetInvestorRelations,
  title = {Alphabet {{Investor Relations}} - {{Investors}}},
  author = {{Alphabet Inc.}},
  year = 2026,
  url = {https://abc.xyz/investor/},
  urldate = {2026-08-19}
}

@online{metaMetaFinancials,
  title = {Meta - {{Financials}}},
  author = {{Meta Platforms, Inc.}},
  year = 2026,
  url = {https://investor.atmeta.com/financials/},
  urldate = {2026-08-19},
  langid = {american}
}

@techreport{pwcIABInternetAdvertising2026,
  title = {{{IAB Internet Advertising Revenue Report}} - {{Full-year}} 2025 Results},
  author = {PwC},
  year = 2026,
  month = apr,
  institution = {PwC},
  urldate = {2026-08-18}
}

@incollection{lomborgDigitalTrackingInfrastructural2023,
  title = {Digital Tracking and Infrastructural Power},
  booktitle = {Handbook of {{Critical Studies}} of {{Artificial Intelligence}}},
  author = {Lomborg, Stine and Helles, Rasmus and Lai, Signe Sophus},
  year = 2023,
  month = nov,
  pages = {354--366},
  publisher = {Edward Elgar Publishing},
  urldate = {2026-08-19},
  chapter = {Handbook of Critical Studies of Artificial Intelligence},
  isbn = {978-1-80392-856-2},
  langid = {english}
}

@article{freitagRealClimateTransformative2021,
  title = {The Real Climate and Transformative Impact of {{ICT}}: {{A}} Critique of Estimates, Trends, and Regulations},
  shorttitle = {The Real Climate and Transformative Impact of {{ICT}}},
  author = {Freitag, Charlotte and Berners-Lee, Mike and Widdicks, Kelly and Knowles, Bran and Blair, Gordon S. and Friday, Adrian},
  date = {2021-09-10},
  journaltitle = {Patterns},
  shortjournal = {Patterns},
  volume = {2},
  number = {9},
  eprint = {34553177},
  eprinttype = {pubmed},
  publisher = {Elsevier},
  issn = {2666-3899},
  doi = {10.1016/j.patter.2021.100340},
  url = {https://www.cell.com/patterns/abstract/S2666-3899(21)00188-4},
  urldate = {2026-06-13},
  langid = {english}
}

\appendix

\section{Initial Paper Titles}
\label{sec:initialpapers}

The list of ten paper titles used to extract query terms from.

\begin{itemize}
    \item An empirical study on the performance and energy costs of ads and analytics in mobile web apps
    \item Carbolytics: An Analysis of the Carbon Costs of Online Tracking
    \item Client-side energy and GHGs assessment of advertising and tracking in the news websites
    \item Energy Conservation with Open Source Ad Blockers
    \item Environmental impact assessment of online advertising
    \item Impact of Ad Blockers on Computer Power Consumption while Web Browsing: A Comparative Analysis
    \item The Effect of Analytics Tools on Energy Consumption of Websites
    \item The Environmental Costs of Surveillance Capitalism: A Case Study of Social Media Platforms
    \item The Hidden Energy Cost of Web Advertising
    \item Users Pay Twice: The Hidden Energy Cost of Web Advertising
\end{itemize}

\section{Search Queries}

Google Scholar search syntax: '|' means OR, a space means AND, parentheses means grouping, and double quotes are use for exact match.

\subsection{Query Terms}
\label{sec:queryterms}

\paragraph{Type of Impact}
\begin{itemize}
    \item (ecological | environmental) (cost | costs | impact | impacts | footprint | sustainability)
    \item (energy | carbon) (cost | costs | consumption | conservation | impact | impacts | footprint | assessment)
    \item power (consumption | assessment)
\end{itemize}

\paragraph{Topic of Focus}
\begin{itemize}
    \item (online | web | website | websites | digital) (ad | ads | analytics | advertising | advertisement | tracking | surveillance)
    \item ("ad blocker" | "ad blockers" | "traffic flows" | "surveillance capitalism")
\end{itemize}

\paragraph{Excluded Terms\\}
-road -urban -cities -gas

\subsection{Final Queries}
\label{sec:finalqueries}
\begin{itemize}
    \item allintitle: (ecological | environmental) (cost | costs | impact | impacts | footprint | sustainability) (online | web | website | websites | digital) (ad | ads | analytics | advertising | advertisement | tracking | surveillance) -road -urban -cities -gas
    \item allintitle: (ecological | environmental) (cost | costs | impact | impacts | footprint | sustainability) ("ad blocker" | "ad blockers" | "traffic flows" | "surveillance capitalism") -road -urban -cities -gas
    \item allintitle: (energy | carbon) (cost | costs | consumption | conservation | impact | impacts | footprint | assessment) (online | web | website | websites | digital) (ad | ads | analytics | advertising | advertisement | tracking | surveillance) -road -urban -cities -gas
    \item allintitle: (energy | carbon) (cost | costs | consumption | conservation | impact | impacts | footprint | assessment) ("ad blocker" | "ad blockers" | "traffic flows" | "surveillance capitalism") -road -urban -cities -gas
    \item allintitle: power (consumption | assessment) (online | web | website | websites | digital) (ad | ads | analytics | advertising | advertisement | tracking | surveillance) -road -urban -cities -gas
    \item allintitle: power (consumption | assessment) ("ad blocker" | "ad blockers" | "traffic flows" | "surveillance capitalism") -road -urban -cities -gas
\end{itemize}

\end{document}